\documentclass[aps,prl,amsmath,twocolumn,amssymb,titlepage,superscriptaddress,showpacs]{revtex4-1}
\usepackage[english]{babel}
\usepackage{graphicx}
\usepackage{transparent}
\usepackage{amsmath}
\usepackage{amssymb}
\usepackage{epstopdf}
\usepackage{amsfonts}
\usepackage{bm}
\usepackage{mathdots}
\usepackage{times}
\usepackage{mathtools}
\usepackage{stmaryrd}
\usepackage{hyperref}
\usepackage{tabularx}
\usepackage{hhline}

\newcommand{\I}{\mathrm{i}}

\newcommand{\ave}[1]{\langle #1 \rangle}

\newcommand{\astcycl}{\mathrlap{\kern0.085em{\circlearrowright}}\ast}
\newcommand{\taucycl}{\mathrlap{\kern0.42em{\bullet}}\circlearrowright}

\begin{document}

\title{Dynamics of photo-doped charge transfer insulators}
\author{Denis Gole\v z}
\affiliation{Department of Physics, University of Fribourg, 1700 Fribourg, Switzerland}
\author{Lewin Boehnke}
\altaffiliation{Now at Crypto Finance AG}
\affiliation{Department of Physics, University of Fribourg, 1700 Fribourg, Switzerland}
\author{Martin Eckstein}
\affiliation{Department of Physics, University of Erlangen-N\"urnberg, 91058 Erlangen, Germany}
\author{Philipp Werner}
\affiliation{Department of Physics, University of Fribourg, 1700 Fribourg, Switzerland}


\begin{abstract}
We study the dynamics of charge-transfer insulators after 
photo-excitation using the three-band Emery model and a 
nonequilibrium extension of Hartree-Fock+EDMFT and GW+EDMFT. 
While the equilibrium properties are accurately reproduced by the Hartree-Fock treatment of the full $p$ bands,
dynamical correlations are essential for a proper description of the photo-doped state. 
The insertion of doublons and holons leads to a renormalization of the charge transfer gap 
and to a substantial 
broadening of the bands. We calculate the time-resolved photoemission spectrum and optical conductivity and 
find qualitative agreement with experiments.  
Our formalism 
enables the realistic description of nonequilibrium phenomena in a large class of charge-transfer insulators, and 
provides a tool to explore the optical manipulation of interaction and correlation effects.  
\end{abstract}

\maketitle

The ability to engineer the microscopic parameters of correlated electron materials 
is important for the understanding of their complex properties and for technological applications. 
An example is the control of superconductivity by pressure, doping or strain \cite{sipos2008,zadik2015,ivashko2018}
under equilibrium conditions. 
Recent 
progress in ultrafast laser techniques 
motivates new strategies which involve nonthermal states. Experiments have demonstrated, e.g., the switching to 
a hidden metallic state in 1T-TaS$_2$ \cite{stojchevska2014}, 
or light-induced enhancement of superconductivity \cite{fausti2011,mitrano2016} and excitonic order \cite{mor2017}. Related are theoretical proposals to modify
the band structure \cite{oka2009,wang2013} or interactions 
by electronic excitations \cite{golez2015,golez2017,tancogne2017} or
phonon driving \cite{knap2016,babadi2017,murakami2017,mazza2017,kennes2017,murakami2017Photo}. 

A simple way to change the properties of correlated
materials is the photo-excitation of charge carriers across a Mott or
charge transfer (CT) gap. 
When
it comes to nonequilibrium simulations most of the theoretical work
related to  Mott insulators has focused on
single-band
models \cite{aoki2014_rev,eckstein2016,eckstein2014a,eckstein2012,eckstein2011a,golez2014,dalconte2015,zala2013},
which miss important aspects of the physics of CT insulators with $p$ and $d$ bands. In these compounds, photo-excitation results in doublon and holon charge carriers of a qualitatively different nature, and depending on the excitation energy one may selectively excite electrons from occupied $p$ or $d$ states. Furthermore,
photo-doping may change the relative position of the $p$ and $d$ bands with a potentially large impact on the low-energy properties.

A paradigmatic class of materials where this plays a role are the cuprates.
The equilibrium properties of cuprates have been investigated with a
broad range of methods including exact diagonalization
\cite{jaklivc2000finite,tohyama2004}, dynamical mean field theory
(DMFT) \cite{comanac2008,medici2009,weber2008} and its cluster
extensions \cite{yin2008}. These studies emphasized the importance of
a multiband description including $p$ orbitals \cite{hansmann2014,ebrahimnejad2014} and the role of spin
\cite{yin2008} and charge fluctuations \cite{werner2015Dynam,werner2016b}. 
Photoexcitation with typical 1.5 eV or 3 eV laser pulses can clearly reveal the multi-band physics, even though the equilibrium properties are partially captured by a single-band model.
In fact, pump-probe studies on cuprate superconductors
have detected photo-induced 
band shifts and modifications in effective masses
\cite{rameau2014,matsuda1994,okamoto2010,okamoto2011,novelli2014,cilento2018},
as well as 
a redistribution of spectral weight
between different characteristic energy scales \cite{mansart2013,giannetti2011}.

In this work, we consider the two-dimensional three-band Emery model \cite{emery1987} 
describing the Cu $d_{x^2-y^2}$ (denoted $d$) and $O$ $p_x$ and $p_y$ orbitals, where the $O$ orbitals lie between Cu ions forming a square lattice with lattice constant $a$. The three terms of the Hamiltonian
$H= H_\text{e}+H_\text{kin}+H_\text{int}$  are
\begin{align}
  H_\text{e}=& \epsilon_d \sum_i n_{i}^d +(\epsilon_d+\Delta_{pd}) \sum_{i,\delta} n_{i}^p , \nonumber\\
  H_\text{kin}=&\sum_{ij\sigma}\sum_{(\alpha,\beta) \in(d,p_x,p_y)} t_{ij}^{\alpha\beta} c_{i\alpha
                 \sigma}^{\dagger} c_{j\beta\sigma}, \nonumber\\
  H_\text{int}=& \sum_{ij}\sum_{(\alpha,\beta) \in(d,p_x,p_y)} V_{ij}^{\alpha\beta} n_i^\alpha n_j^\beta,\nonumber
\end{align}
with $V_{ij}^{\alpha\beta}=U_{dd}$ if $i=j$ and $\alpha=\beta=d$,
and $V_{ij}^{\alpha\beta}=\frac{1}{2}U_{dp}$ for nearest-neighbor $d$ and $p$ orbitals. 
 The crystal field splitting $\Delta_{dp}$ determines the difference
between the on-site energy for the $d$ orbital, $\epsilon_d$, and that for the $p$ orbitals, $\epsilon_d+\Delta_{dp}.$ We
denote the nearest neighbor hopping between the $d$ and $p_x$, $p_y$ orbitals by $t^{dp}$ and between the $p_x$
and $p_y$ orbitals by $t^{pp}$, while $-t^{\alpha\alpha}_{ii}=\mu$ is
the chemical potential. 

We solve this lattice problem using the fully self-consistent GW+EDMFT  \cite{biermann2003,ayral2013,huang2014,golez2017}
and Hartree-Fock (HF)+EDMFT 
methods, which are
based on extended dynamical mean field theory
(EDMFT) \cite{sun2002,ayral2013,huang2014,golez2015}, and 
a non-crossing approximation impurity solver \cite{golez2015}. Since the local
interactions in the $d$ orbital are stronger than in the $p$ orbitals,
we
restrict the correlated subspace (EDMFT treatment) to the $d$-orbital,
while the $p$ orbitals are treated at the HF or $GW$
level. This approach requires a downfolding from the
three-orbital space of the full model to the $d$ orbital correlated
subspace, which has to be implemented 
at the level of the electronic Green's function and the (bosonic)
screened interaction.

The dynamics of the system is described in terms of the momentum and
orbital resolved electronic Green's function
$ G_{k\sigma}^{\alpha\beta}(t,t')= -\I \ave{\mathcal{T}_\mathcal{C}
  c_{k\alpha\sigma}(t) c_{k\beta\sigma}^{\dagger}(t')} $ and the
charge correlation function
$\chi_{q}^{\alpha\beta}(t,t')=-i\ave{ T_\mathcal{C} \tilde
  n_{q\alpha}(t) \tilde n_{-q\beta}(t')}$, which determines the inverse
of the orbital-resolved dielectric function
$[\varepsilon^{-1}_{q}]=I+v_{q}\ast
\chi_{q}$ and the screened interaction
$W_{q}=[\varepsilon^{-1}_{q}] \ast
v_q$, where $v_q$ is the Fourier transform
of the bare interaction. 
The mapping of the lattice problem to
an impurity problem and the downfolding procedure lead to a 
retarded density-density interaction $\mathcal{U}(t,t')$ on the  impurity ($d$ orbital),
which is related to the screened interaction
by $W_\text{loc}^{dd}=\mathcal{U}+\mathcal{U}*\chi_\text{imp}*\mathcal{U}.$

\begin{figure}[t]
\includegraphics{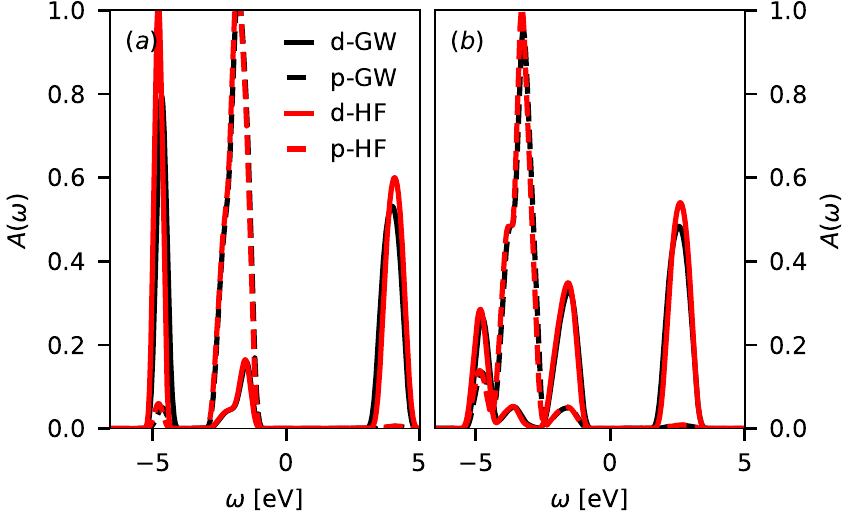}\\
\caption{ Orbitally resolved spectral function 
  obtained in the GW+EDMFT (black) and HF+EDMFT (red) approximation
  for the CT (a)
  and LCO (b) set-ups. 
  The full (dashed) lines represent the
  $d$ ($p_x$ and $p_y$) orbitals.}
  \label{Fig:Spectral}
\end{figure}

The electromagnetic field leads to an 
acceleration of the charge carriers as well as dipolar transitions. The 
gauge invariant 
description of both effects is given
by the modified hopping term \cite{loudon2000, golezFUT} 
\begin{align}
H_\text{kin}=&\sum_{ij\sigma,(\alpha,\beta)\atop \in(d,p_x,p_y)} (t_{ij}^{\alpha\beta}+ \vec E \cdot \vec
  D_{ij}^{\alpha\beta} ) e^{-\I
    \varphi_{ij}} c_{i\alpha
                 \sigma}^{\dagger} c_{j\beta\sigma},
\end{align}
where $\varphi_{ij}=e/\hbar \int_{R_i}^{R_j} d\vec r \vec A(\vec r,t)$ is the Peierls phase and $\vec A(\vec r,t)$ is the vector potential, which is
related to the electric field by $\vec E(\vec r,t)=-\partial_t \vec A(\vec \rho,t)$.
We assume that $\vec A(t)$ and $\vec E(t)$ are homogeneous in space. 
For a quantitative description of the photo-excitation in
multiband materials the dipolar matrix elements will have to be obtained from ab-initio calculations.
In the following simulations, we use 
nearest neighbor $d$ to $p$ dipolar matrix elements $D \equiv D_{(i,\{0,0\})(i,\{a/2,0\})}^{p_xd}=-D_{(i,\{0,0\})(i,\{a/2,0\})}^{p_yd}=0.3\textup{ e\AA}$. We have checked that the results do not qualitatively dependent on the value of $D$. 

We will study two characteristic set-ups previously
considered in the literature: i) a charge transfer insulator close to
the atomic limit, where a strong Coulomb repulsion opens a large Mott
gap and narrow $p$ bands lie well separated between the lower and
the upper Hubbard bands (``CT case"), and ii) the parameter regime relevant for
La$_2$CuO$_4$, as obtained from LDA calculations and the constrained
RPA downfolding procedure to the low energy space composed of the $d$
and $p$ orbitals \cite{werner2015Dynam} (``LCO case").  In this
set-up the $p$ bands lie in the same energy range as the lower Hubbard
band. We parametrise the tight-binding Hamiltonian following
Ref.~\onlinecite{hansmann2014}.

\begin{figure}[t]
\includegraphics{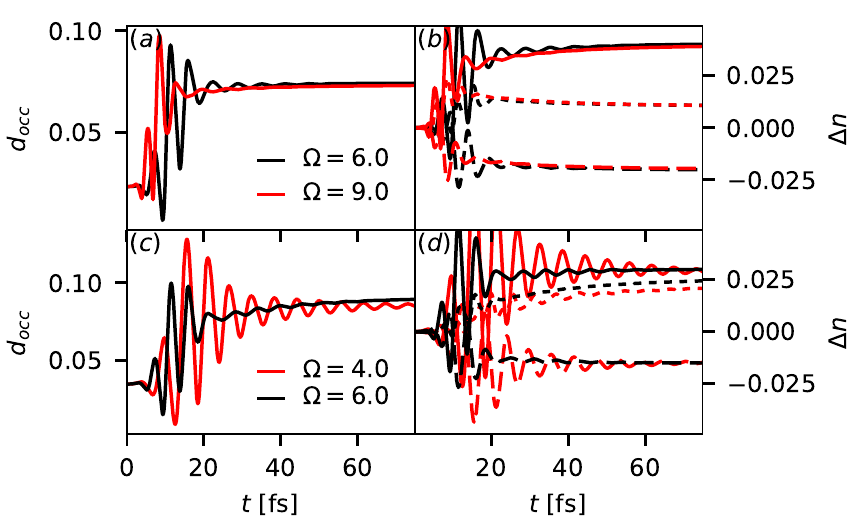}\\
\caption{Dynamics after photo-excitation for the CT 
(top panels) 
and LCO
(bottom panels)
set-up.  The left panels show the change of the 
  double occupancy $d_\text{occ}$, while the right panels plot the change in the
  occupation of the $d$ (full line) and $p_x$ $(p_y)$ (dashed line) orbitals. Dotted lines show the 
  density of holes in the lower Hubbard band.
  }
\label{Fig:Excitation}
\end{figure}

In Fig.~\ref{Fig:Spectral} we present the orbitally 
resolved local spectral functions
$A_{\alpha}(\omega)=-\frac{1}{\pi}\text{Im}[G_{\text{loc}}^{\alpha,\alpha}(\omega)],$
for $\alpha=d,p_x,p_y$. In all calculations we
set the inverse temperature to $\beta=5.0$, which is above the
N\'{e}el temperature. 
In the CT case (Fig.~\ref{Fig:Spectral}(a)), with parameters
$U_{dd}=8.0$ eV, $U_{dp}=2.0$ eV, $t_{dp}=0.4$ eV, $t_{dd}=-0.1$ eV, $t_{pp}=0.15$ eV, $\Delta_{pd}=-2.0$ eV, 
the lower Hubbard band is well separated from the $p$ bands. 
Due to the hopping between the $d$ and $p$
orbitals there is a hybridisation 
which pushes the lower Hubbard band 
further down in energy. In the example relevant for La$_2$CuO$_4$ (Fig.~\ref{Fig:Spectral}(b)), 
with parameters $U_{dd}=5.0$ eV, $U_{dp}=2.0$ eV, $t_{dp}=0.5$ eV, $t_{dd}=-0.1$ eV, $t_{pp}=0.15$ eV, $\Delta_{pd}=-3.5$ eV,  
the spectral function below the Fermi level is split into three distinct peaks:
a peak at $\omega=-3.3$ eV of predominantly $p$ orbital character, the 
bonding band corresponding to the Zhang-Rice singlet around $\omega=-1.6$ eV, and the
anti-bonding band pushed to lower energy, with a center at
$\omega=-4.9$ eV.  
In equilibrium, the GW+EDMFT and HF+EDMFT
spectra are almost indistinguishable, which shows that static correlations are sufficient for the treatment
of the fully occupied $p$ orbitals. We also note that  
the gap suppresses the 
effect of screening \cite{ayral2017}, so that the $GW$ self-energy and polarization
results in only a tiny reduction of the effective static interaction (and hence of the gap):
$\mathcal{U}(\omega=0)-U_{dd}\approx -0.03$ eV for the CT and $-0.05$ eV for the LCO set-up.

We now turn to the photo-excitation and the subsequent relaxation.
A short pulse
$E(t) = E_0e^{-4.6(t-t_0)^2/t_0^2} \sin(\Omega(t-t_0))$ polarized along the (11) direction creates
doubly occupied $d$ orbitals as evidenced by an enhanced 
population of the upper Hubbard band, see Fig.~\ref{Fig:Excitation}. The width of the pulse
$t_0 = 2\pi n/\Omega$ is chosen such that the envelope accommodates
$n = 2$ cycles. Due to the mixed nature of the states
below the Fermi level one may expect that the ratio between the excited electrons
originating from the $p$ or $d$ orbital depends on the frequency of
the pulse. 
In the following we will choose frequencies corresponding to
a photo-excitation from the characteristic features in the occupied part of the spectrum 
to the upper Hubbard band and adjust the excitation strength
of the pulse $E_0$ at each given frequency such that the number of 
photo-doped 
doublons and holons 
is approximately $5\%.$

In Fig.~\ref{Fig:Excitation}(a) we present the time evolution of the
double occupancy in the CT set-up for
two different frequencies corresponding to the transitions from the
$p$ band ($\Omega=6, E_0=0.37$) and the lower Hubbard band
($\Omega=9, E_0=0.87$) to the upper Hubbard band. Due to the large gap size the
number of doubly occupied sites changes only very slowly after the
photo-excitation
\cite{sensarma2010,eckstein2011a,zala2014Recomb,zala2013}. 
The change in the occupation of the orbitals (Fig.~\ref{Fig:Excitation}(b)) shows
that charge is transferred from the $p$ to the $d$ orbital 
for both excitations. 
The dotted lines in Fig.~\ref{Fig:Excitation}(b) indicate the density of holes in the lower Hubbard band, $\Delta d_\text{occ}-2\Delta n_p$. The final occupations are remarkably independent of the excitation frequency. The spectrally resolved occupation (not shown) indicates that after an ultra-fast redistribution of occupation between the region of the lower Hubbard band and the $p$-band the holes predominantly reside in the $p$-band. This inter-band decay is only described within the GW+DMFT formalism and not by HF+DMFT, so that its likely origin is the strong coupling of electrons to charge fluctuations. The response is similar to the single band extended Hubbard
model deep in the Mott phase, where the doublons quickly relax to the
lower edge of the upper Hubbard band via the emission of plasmonic
excitations \cite{golez2015}. 
This observation shows that the correct description of the photo-excited state already a few fs after the excitation, and of orbital selectivity, 
requires an accurate
treatment of the dynamic charge fluctuations. 

In the LCO case we excite
electrons to the upper Hubbard band either from the ZR singlet
($\Omega=4, E_0=0.54$) or from the band with dominant $p$ character 
($\Omega=6.0, E_0=0.31$). Due to the smaller gap
size, impact-ionization processes become important and lead to a more
rapid increase of the double occupancy after the photo-excitation
\cite{werner2014}. For a fixed number of doubly occupied sites (Fig.~\ref{Fig:Excitation}(d)) there is
a larger percentage of holes in the $d$
orbitals than for the CT insulator~(dashed lines). The holes predominantly occupy the ZR band as seen in the greater
component of the orbitally resolved spectral function (not shown). Apart from strong initial oscillation due to the
$p$-$d$ hybridization the final occupation is again remarkably independent of $\Omega$.

\begin{figure}[t]
\includegraphics{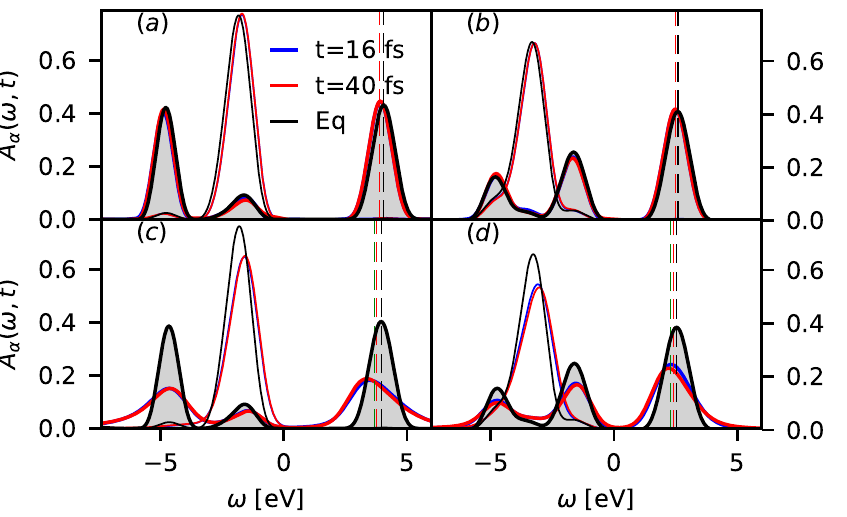}\\
\caption{Time evolution of the spectral function $A_{\alpha}(\omega,t)$ within HF+EDMFT ((a) and (b)) and
  GW+EDMFT ((c) and (d)) for the CT (left panels) and LCO (right panels) set-ups, 
  both excited with the pulse frequency $\Omega=6.0$. 
  The black lines show the equilibrium spectra.  Thick shaded (thin) lines represent the $d$ ($p_x$ and $p_y$)
  orbitals. The black vertical line marks the peak position in 
  equilibrium, the red line indicates the shift of the band due
  to the reduced static Hartree interaction and the green line corresponds to the 
  combined effect of Hartree shift and
  static
  reduction of the effective impurity interaction
  $\mathcal{U}(\omega=0).$
  }
\label{Fig:tPES}
\end{figure}

We next discuss the time evolution of the spectrum after the
photo-excitation by analyzing the partial Fourier
transform of the orbitally resolved spectral function
$A_{\alpha}(\omega,t)=-(1/\pi)\text{Im}[\int_{t}^{t+t_\text{cut}} dt'
e^{\mathrm{i} \omega (t'-t) } G^{R}_{\alpha}(t',t)]$, with
$t_{\text{cut}}=8$ and $\alpha\in\{d,p_x,p_y\}$. Again we will
compare the GW+EDMFT and HF+EDMFT approximations in order
to address the role of nonlocal charge fluctuations. Due to the Coulomb interaction $U_{dp}$ between the 
holes in the $p$ orbitals and the doublons in the $d$ orbital
there is an 
almost instantaneous reduction of the gap between the
$p$ band and the upper Hubbard band, see Fig.~\ref{Fig:tPES}.
It originates from the static Hartree-Fock attraction and is 
determined by the change in the local orbital occupancy \cite{cilento2018}
$\Delta \Sigma^\text{H}_{dd}=(U_{dd}-2U_{dp})\Delta n_{d}$, 
where
the factor of $2$ originates from the number of nearest-neighbor $p$ orbitals and
we have used the conservation of the total charge
$\Delta n_d =-(\Delta n_{p_x}+\Delta n_{p_y}).$ The 
contribution of this effect to the shrinking gap is indicated 
in Fig.~\ref{Fig:tPES} by the vertical red line, $\Delta
\Sigma^\text{H}_{dd}=-0.15$ $(-0.17)$ eV for the CT (LCO) case. 

Going beyond the HF description, the inclusion of nonlocal charge
fluctuations leads to: a) a further reduction of the gap size and a \textit{band shift},
and
 b) \textit{a substantial broadening} of the spectra (in particular for the $d$
orbitals) due to a strong electron-plasmon coupling (compare the GW+EDMFT and HF+EDMFT results in
Fig.~\ref{Fig:tPES}). 
Both effects are a consequence of the photo-induced
changes in the screening. 
The additional gap shrinking can be quantified by the reduced static interaction,
$\mathcal{U}(\omega=0)-U_{dd}\approx-0.14$ eV for both set-ups, which is indicated by the green vertical line in Fig.~\ref{Fig:tPES}. 
One can see that the full reduction of the gap cannot
be described by these static contributions alone and must be attributed to 
dynamical screening and doping effects. By analyzing the momentum-dependent spectrum we confirmed 
that the broadening of the local (momentum-averaged) spectrum originates mainly from a strong increase in the
linewidth due to the electron-plasmon coupling and to a much lesser extent from a modification of the effective velocities.
 Comparing Fig.~\ref{Fig:tPES} with Fig.~\ref{Fig:Spectral} we conclude that the treatment of
both dynamic screening and the feedback of the  nonlocal
charge fluctuations is qualitatively important for the
description of the photo-excited state, while these
effects are negligible in equilibrium \footnote{We
have checked that for high-frequency excitations the shape of the
spectra after the pulse is essentially the same.}.

The dynamics of the $p$ band exhibits a remarkable
resemblance with recent time-resolved ARPES data, which show a
non-thermal broadening of the $2p_{\pi}$ band in the antinodal
direction of optimally doped Y-Bi2212 \cite{cilento2018}. The
interpretation of Ref.~\onlinecite{cilento2018}, which attributed the
non-thermal changes in the experimental spectra to nonlocal charge
fluctuations, is fully consistent with our results.

\begin{figure}[t]
\includegraphics{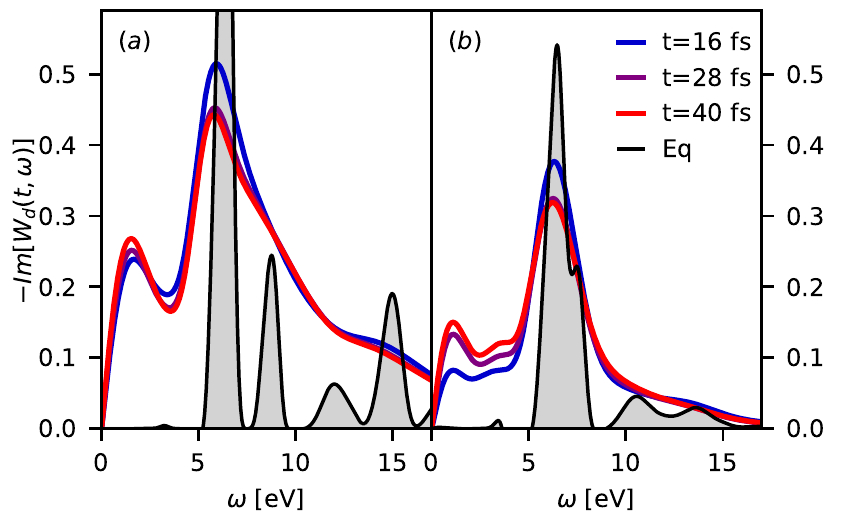}
\caption{Time evolution of the screened interaction $W_{d}(t,\omega)$ for the CT~(a) and LCO~(b) set-up, both excited with frequency $\Omega=6.0$.  The black shaded lines represent the
  equilibrium spectrum. 
  }
\label{Fig:W}
\end{figure}
The information about the
time-dependent changes in the screening properties of the $d$ orbital is
contained in the screened interaction $W_{d}(\omega,t)$,
 see Fig.~\ref{Fig:W}.  
The equilibrium screening modes for
the charge-transfer insulator include i) a peak at $\omega_{CT} \approx 6$
corresponding to particle-hole excitations from the $p$ band to
the upper Hubbard band, and ii) a peak at $\omega_{H}\approx 8.5$ matching
the excitations from the lower to the upper Hubbard band, and higher
order excitations [see also 
Fig.~\ref{Fig:Spectral}(a)].
After the photo-excitation these features are smeared
out and a continuum of screening modes appears at low energies. 
These are associated with charge excitations within the photo-doped bands and  
result in an additional screening 
\cite{golez2015}.
Within EDMFT one can define an effective coupling of the electrons and the charge fluctuations
from the integral over $\text{Im}\mathcal{U}(\omega)/\omega$ (or, equivalently, the reduction of $\text{Re}\mathcal{U}(\omega=0)$) \cite{ayral2013,golez2015,golez2017}.
The coupling strength to the photo-induced charge fluctuations at $\omega \lesssim 6$,
$\lambda_\text{ind}= 2\int_0^{6}
  d\omega (-\text{Im}[\mathcal{U(\omega)}]/\omega) \approx 0.16$,  
  corresponds to a rather strong coupling, which explains the
substantial broadening of the spectra 
in Fig.~\ref{Fig:tPES}. The dynamics of  
$W_d(t,\omega)$
in the LCO case 
is qualitatively similar except for an additional peak 
at $\omega\approx3.5$, which originates from the
charge excitations
between the Zhang-Rice singlet and the lower edge of the upper Hubbard
band. The effective electron-plasmon coupling
$\lambda_\text{ind}\approx0.17$ is comparable to the CT case.

Finally, we discuss the signature of the
band-gap renormalization in the optical conductivity. We
explicitly simulate a probe pulse along the (11) direction and extract the photo-induced
current as the difference in the current with and without a probe pulse,
$j_\text{probe}=j_\text{pump+probe}-j_\text{pump}.$ For a weak probe pulse the optical
conductivity can be evaluated as the ratio $\sigma(\omega,t_p)=j(\omega,t_p)/E(\omega,t_p),$ where
$X(\omega,t_p)=\int_0^{t_\text{cut}} ds X(t+s) e^{-\I \omega s}$ is the Fourier transform of $X=j,E,$ and
$t_p$ is the center of the probe pulse. 
This procedure avoids the calculation of the current-current correlation function
including vertex corrections. We apply both the pump and probe pulses in the (11) direction.

\begin{figure}[t]
\includegraphics{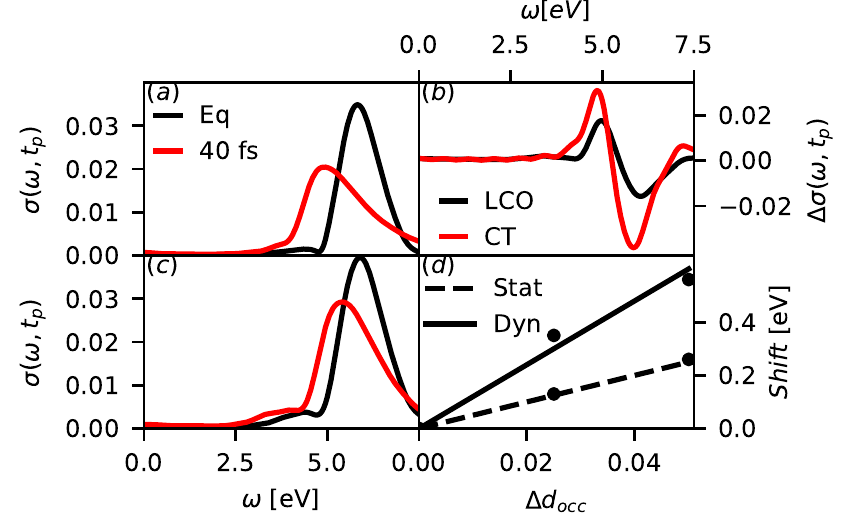}\\
\caption{Time-dependent optical conductivity in the (11) direction $\sigma(\omega,t_p)$ for the CT (a) and LCO (c) set-up. The black lines represent the equilibrium data,
where we used an additional damping factor $\eta=0.25$ to smoothen artifacts of the Fourier transform due to the finite time window. (b) The difference from the equilibrium result
  $\Delta\sigma(t_p,\omega)=\sigma(t_p,\omega)-\sigma^\text{eq}(\omega)$ for the delayed probe pulse $t_p=40$ fs in both set-up. (d) The  gap size renormalization at the position of the half-maximum in the optical conductivity (Dyn) as a function of the photo-excited doublons and comparison with the renormalization expected from the static shift (Stat) $\Sigma_{H}+\mathcal{U}(\omega=0)$. 
}
\label{Fig:optical}
\end{figure}

The optical conductivity in equilibrium (black lines) and for 
$t_p=40$ fs after the pump pulse is presented in
Fig.~\ref{Fig:optical}(a,c). 
The equilibrium optical
conductivity for both cases exhibits a main peak corresponding to
excitations from the $p$ band to the upper Hubbard band. In the
LCO case, see Fig.~\ref{Fig:optical}(c), the small
peak at $\omega=4.4$ eV corresponds to excitations from the ZR
singlet to the upper Hubbard band.

After the photo-excitation the optical conductivity shows a clear
shift toward lower energies, originating from the HF shift
and the enhanced screening. 
In order to
highlight this evolution we also plot the change of the optical conductivity 
from the equilibrium value (Fig.~\ref{Fig:optical}(b)) as it is usually
done in the experimental literature \cite{novelli2014}. Indeed, a 
characteristic shift of the excitation gap to lower energies has been 
reported in a number of pump-probe experiments on
cuprates and other charge transfer insulators, see for instance
Refs.~\onlinecite{matsuda1994,giannetti2016,novelli2014,okamoto2011,okamoto2010}. 
The comparison of the shift to the static screening contribution (line labeled ``Stat" in Fig.~\ref{Fig:optical}(d))
shows that more than 50\% of the gap renormalization is due to the broadening of the bands.
A recent experimental study of photo-doped La$_2$CuO$_4$ \cite{novelli2014} reports an 
even somewhat larger shift than predicted by the line labeled ``Dyn" in Fig.~\ref{Fig:optical}(d).
This suggests that dynamical effects, including the substantial broadening of the bands in the photo-doped state, are essential for a quantitative description of photo-doped cuprates.

In conclusion, we have studied the pump-probe dynamics of 
the 3-band Emery model, relevant for a large family of 
charge-transfer insulators, using nonequilibrium GW+EDMFT and HF+EDMFT. The electric field pulse 
transfers charge from the $p$ bands to the upper Hubbard band, 
which results in relative band shifts. Similar band shifts have been observed in a recent TDDFT+U study of NiO \cite{tancogne2017}, while the present study shows that the dynamical screening strongly enhances this effect. The strong plasmon coupling in the photo-doped state leads to a substantial broadening of the $d$ and $p$ bands, which implies that 
dynamical correlation effects (beyond HF) are essential for the description of the ligand bands, in contrast to equilibrium.

The electronic structure of the Zhang-Rice singlet determines the low-energy magnetic
properties of the system. The photo-induced shift may have profound
consequences on the magnetic correlations in these systems, which makes  the investigation of 
photo-induced changes in the magnetic correlations an interesting topic for
future studies. On the methodological side, our work represents a crucial step toward
the \textit{ab-initio} description of strongly correlated systems out of
equilibrium, where the material-specific input is obtained via a
multi-tier approach analogous to the scheme recently demonstrated for
equilibrium systems in Refs.~\cite{werner2015Dynam,boehnke2016,nilsson2017}.

\acknowledgements 

We thank Y. Murakami and A.~J. Millis for useful discussions,
and M.~Sch\"uler for numerous advises on computational issues. The
calculations have been performed on the Beo04 and Beo05 clusters at the
University of Fribourg.  DG and PW acknowledge support from SNSF Grant No. 200021\_165539 and
ERC Consolidator Grant No. 724103. ME acknowledges financial support from ERC Starting Grant No. 716648.

\bibliography{../../BibTex/tdmft,../../BibTex/Books,../../BibTex/Polarons}

\end{document}